\documentstyle[aps,preprint,prl,epsfig,epsf]{revtex}

\tightenlines
\begin{document}

\title{Role of the experimental filter in obtaining the Arrhenius plot in
multi-fragmentation reactions.}
\author{O. Tirel$^1$, G. Auger$^1$, R. Nebauer$^{2,+}$, J. Aichelin$^2$}
\date{\today}
\maketitle

\noindent 1 GANIL, CEA/DSM CNRS-IN2P3, BP 5027, 14076 Caen Cedex 05, France \\
2 SUBATECH, Universit\'e de Nantes, EMN, IN2P3/CNRS, 4, rue Alfred Kastler, 44070 Nantes Cedex 03, France \\
+ Universit\"at Rostock, Germany

\section{abstract}
Recently it has been argued that the linear relation between the transverse
energy and the apparent probability to emit a fragment proves that the total
system is in thermal equilibrium. It is shown, for the specific reaction
Xe+Sn at 50 A.MeV, that the same behavior is obtained in the context of
Quantum Molecular Dynamical without invoking the idea of equilibrium. The
linear dependance is shown to be a detector effect.

\clearpage

The process of multiple emission of intermediate mass fragments
(I.M.F), so-called multi-fragmentation, is one of the central issues in
heavy ion reactions at intermediate energies : Is the multi-fragmentation a
thermal or a dynamical process or a mixing of both?\newline
Some recent papers try to demonstrate that production of I.M.F. can be
essentially attributed to dynamical effects \cite{toke,tirel,lukasik}, at
the same time other authors claim that multi-fragmentation is a thermal
process \cite{marie,botvina}. Recently it has been conjectured by L.Moretto
that the observation of regularities in I.M.F production rates, viewed as
functions of "temperature" have the form of an Arrhenius plot \cite
{moreto,tso,wieloch}. This finding, well known from chemical process in
equilibrium, is for the authors \cite{moreto,tso} a proof that
multi-fragmentation is indeed a thermal process.\newline
The analysis is based on the fact that the probability $p$ 
of one-fragment emission is independent of the emission of the other
fragments, so the probability $p$ for the elementary emission
of I.M.F is related to the binomial law :

\begin{equation}
P_{n}^{m}(p)=\frac{m!}{n!(m-n)!}p^{n}(1-p)^{(m-n)}  \label{binomial}
\end{equation}

 where $m$ is the number of chances to emit a
fragment by the system and $n$ is the multiplicity of fragment. 

The process is considered as thermal, if $log(1/p)$ versus the inverse temperature $(1/T)$ is linear. In the ref.\cite{moreto}
it is assumed that the transverse energy ($E_{t}=\sum_{i}E_{i}sin^{2}(\theta
_{i})$, where $E_{i}$ is the kinetic energy of all charged particles, and $%
\theta_{i}$ the polar angle) is proportional to the excitation energy $E^{*}$
which in the Fermi gas model is proportional to the temperature squared ($
E^{*}\propto T^{2}$), so that the square root of $E_{t}$ is proportional to
the temperature ($\sqrt{E_{t}}\propto T$). \newline

 Various point of this analysis can be critisized :

\begin{itemize}
\item  the transverse energy contains in part the transverse
energy of the I.M.F. and because of that, the multiplicity $n$
\ and the transverse energy are correlated \cite{toke2,tsang,wieloch}. In
reference \cite{tso} and in some recent publications \cite{moreto2,beaulieu}
the authors claim that the Arrhenius law seen is not due to such
correlations. 

\item  It is based on the hypothesis that the detection efficiency, for a 4$
\pi $ detector, is independent of the reaction mechanism i.e of the impact
parameter and on the total multiplicity.

\item  All the fragments are not emitted by a unique source. In this case it is difficult to believe that a unique temperature can be
achieved for all the sources. 
\end{itemize}

 The aim of this paper is to show that the Arrhenius law can be
obtained with a model without the idea of thermalization and also to
demonstrate the effect of the detector efficiency. In the first part,
dynamical calculations on the total range of impact parameters is used, then
the relation between $log(1/p)$ and $E_{t}$ is plotted with events filtered
by the response of one detector. In the second part a simplified filter is
used in order to explain the role of the detector in obtaining the Arrhenius
law.

The Quantum-Molecular Dynamic model (QMD)  \cite{gossiaux} shows a good agreement with experimental heavy-ions reaction
data at high incident energies. We have compared the data of the Xe+Sn
reaction at 50 A.MeV obtained with the INDRA detector \cite{pouthas} with the
dynamical QMD code \cite{aichelin} For this we have calculated 63000 events
with an impact parameter between 0 and 12 fm in order to cover the complete
reaction cross section. We find a very good agreement for a large part of
the reaction cross section both for kinematical (velocity of I.M.F. and
light particles ...etc.), and statistical variables (fragments multiplicity,
charge distribution...etc.)\cite{tirel}. It was thus concluded that the
reaction mechanism process is mainly binary and that the IMF's seem to be
produced dynamically and not by thermalized sources. The necessary
assumption to observe the Arrhenius law is normally, in these simulations, not
satisfied.

On the contrary, in recent publication, A.Wieloch et al. found the linear
dependance of $log{1}/{p}$ versus ${1}/{\sqrt{E_{t}}}$ in the data of 
 Xe+Sn at 50 A.MeV. We have therefore used the QMD
events to construct an Arrhenius plot. All the events were filtered with
the code simulating the total response  (energy threshold,
geometrical aperture, calibration and identification as in the experiment) 
of the INDRA detector  \cite{pouthas}.

 For each event we calculate the total transverse energy ($E_{t}$
) and for each bin in $E_{t}$ we extract the value of $p$ and $m$ from the
prescription of ref. \cite{moreto} with the following formulae : 
\begin{equation}
p=1-\frac{\sigma ^{2}}{<N_{IMF}>},  \label{proba}
\end{equation}
\begin{equation}
m=\frac{<N_{IMF}>}{p},  \label{m}
\end{equation}

 where $N_{IMF}$ is the mean value, and $\sigma ^{2}$ the
variance of the I.M.F multiplicity distribution.

 For the determination of $p$ and $m$ it is clear that only "emitted fragments" must be taken into
account and that the residual nucleus must be excluded. In every such
analysis, special care is therefore taken to treat the detection of
projectile and target residue. In ref \cite{moreto} due to experimental
thresholds the projectile and target residues are never detected and the
charge of I.M.F's is taken to be between 3 and 20 (essentially to eliminate the
fission process, certainly always present in this reaction). In ref. \cite
{wieloch} they do not take into account the projectile residue as an I.M.F
even if its charge is in this interval ($3\leq Z\leq 20)$  and
they assume that the target residue is always lost due to the energy
thresholds. In the present case, we take as IMF's all fragments with a charge
greater than 3, and because of the binary character of the reaction we do
not take into account, in our I.M.F multiplicity, the projectile residue
that it is always detected and, as is it done in the reference \cite{wieloch}
, we assume that the target residue is supposed to be undetected. This
correspond, in this case, to eliminate one fragment from each filtered event.

\begin{figure}[htbp]
  \begin{center}
    \leavevmode
     \epsfxsize=15cm
     \epsfbox{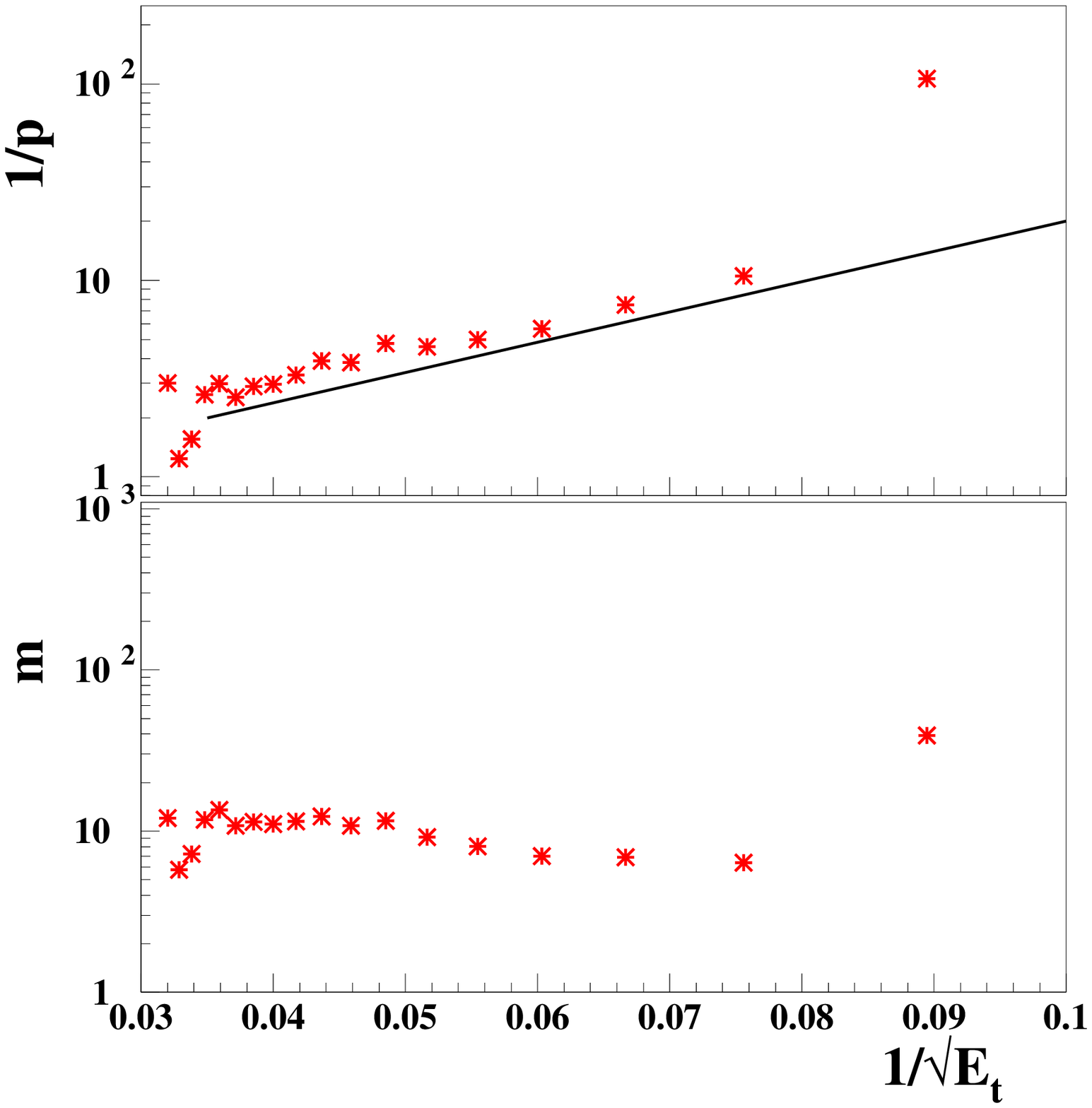}
      \caption{ }
    \label{fig:qmdf}
  \end{center}
\end{figure}

The results of $\log({1}/{p})$ as a function of ${1}/{\sqrt{E_t}}$ (see figure \ref{fig:qmdf}) looks like the
linear Arrhenius law if we do not take into account the hight energy point
for which the variance has no signification, and the first low energies
points for which the statistic is poor (due to the cross section for the low
impact parameters). In the same way, $m$ is rather constant as it is
expected. This entails two questions : first, if the linear Arrhenius plot
is a characteristic of a thermal process why is it found in a
non-thermodynamical calculation ? Second, what is exactly the bias of the
detection system and of the reaction mechanism in obtaining this behavior?

In order to understand the detection effects, an analysis
without filtering the events was performed. In this case, the efficiency is
100\% and only the IMF's are taken into account : the projectile and target
residues are systematicelly removed event by event due to the binary
caracter of the event on the total range of impact parameter.

\begin{figure}[htbp]
  \begin{center}
    \leavevmode
     \epsfxsize=15cm
     \epsfbox{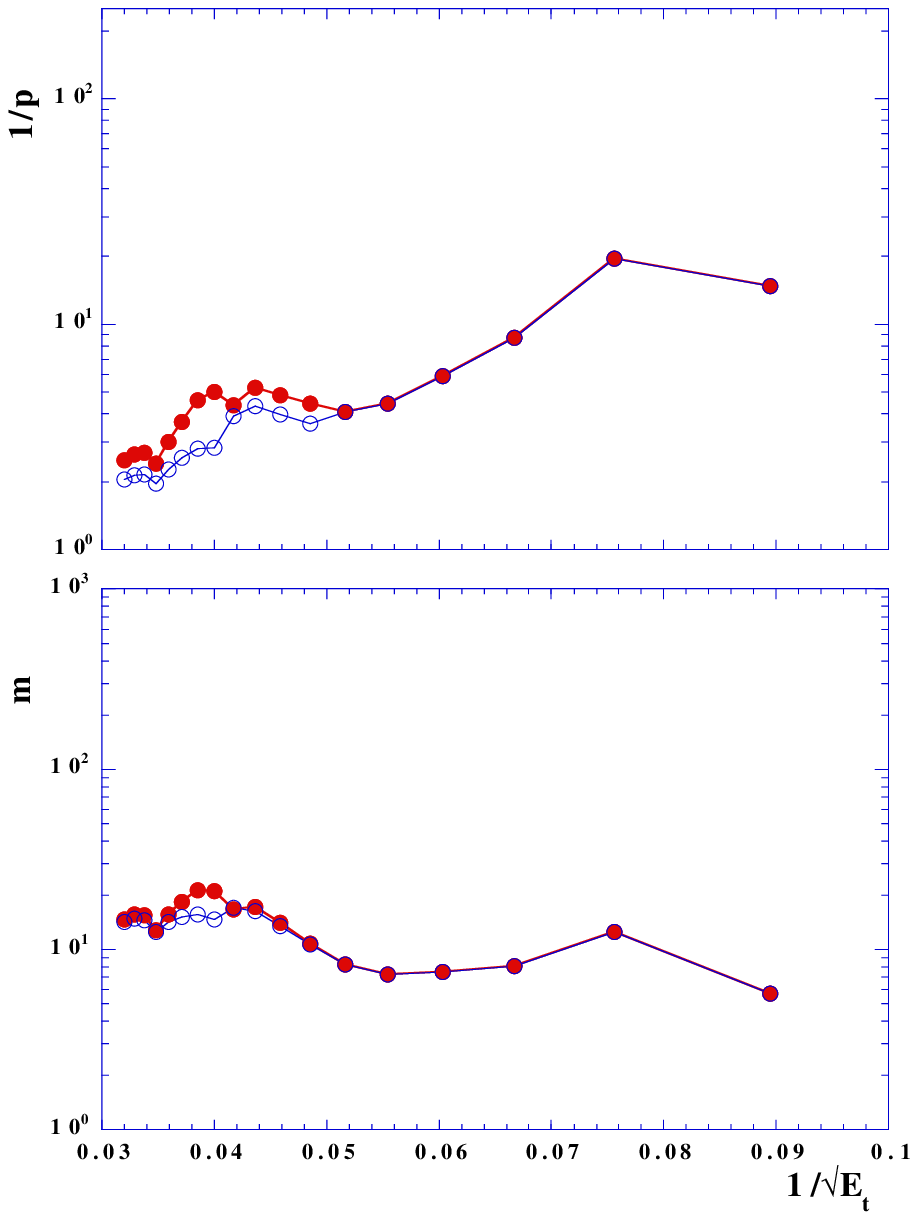}
      \caption{ }
    \label{fig:pm}
  \end{center}
\end{figure}

 Figure \ref{fig:pm} (see full circle) shows that, with
the unfiltered events, the linear aspect of the Arrhenius plot is not
observed anymore. The deviations occur essentially for the semi-peripheral
events with $400<E_{t}(MeV)<600$ which correspond to impact parameter in the
range of $4.5$ $<b(fm)<7.5$ for this reaction in the QMD model.

With respect to figure 1, this different behaviour can only
come from a detector effect. Indeed, for the most peripheral collisions, the
quasi-projectile is always detected while the quasi-target energy is below
the detector threshold. It is therefore normal, taking into account the
behaviour of the detector, to substract, in the $ E_t<400$ region, two fragments (i.e. the quasi-projectile and the quasi-target)
from the distribution of the multiplicity of fragments. 

For the more central collisions, the quasi-target has enough
kinetic energy to be detected.Therefore, to have considered that is was not
detected in constructing figure \ref{fig:qmdf}, corresponds to having substracted not two
but only one fragment. 

In the intermediate region, the number of fragments substracted
is varied smoothly, and statistically, between the two limiting values. 

The experimental construction of the $log(1/p)$ versus $E_t$ curve is
totally equivalent. Indeed, restricting IMFs to have charges between 3 and
20 and substracting only one fragment if it is the "largest" is equivalent
to : 

\begin{itemize}
\item  In the low transverse energy to substract two fragments
to the total multiplicity,the quasi-projectile whose charge is greater than
20 and the quasi-target because it is not detected. 

\item  In the high transverse energy region, the quasi
projectile and the quasi-target are, in most cases, detected with a charge
less than 25 (see the velocity distribution in the laboratory of the two
heaviest fragments in the Xe+Sn reaction \cite{benlliure}.The elimination of
only the heaviest fragment is equivalent to substracting only one fragment. 

\item For the intermediate energy region, a progressive
transition between these two cases is the probable scenario. We have
filtered QMD events with the prescription summarized in table \ref
{tab:coupure} and the results are shown on figure \ref{fig:pm} (see
open circle). 
\end{itemize}

\begin{table}[htbp]
\begin{center}
\begin{tabular}{|c|c|c|c|}

$E_{t}$ (MeV) & 0 \hspace*{3cm} 400  & 400 \hspace*{3cm} 600 & 600 \hspace*{3cm} $\infty$ \\
\hline 
b (fm)        & 12 \hspace*{3cm} 7.5 & 7.5 \hspace*{3cm} 4.5 & 4.5 \hspace*{3cm} 0 \\
\hline
$N_{IMF}$ =   & $N_{fragment}-2$     & $N_{fragment}- [1;2]$ & $N_{fragment}-1$

\end{tabular}
\end{center}
\caption{Number of fragments taken into account for the different domain in $%
E_{t}$. For the range $400<E_{t}(MeV)<600$ we have taken $N_{fragment}-1$
for $E_{t} = 400 MeV$ and $N_{fragment}-2$ for $E_{t} = 600 MeV$, and a
statically linear prescription in between}
\label{tab:coupure}
\end{table}

The function $log(1/p)$ versus $1/\sqrt{E_t}$ become smoother,
i.e the simple filter used reduces the effects and the results look like the
"Arrhenius law". It seems that the filter cancels the former effects in the
region of $E_{t}$. Of course, the real effects of the 4$\pi $ detector is
more complex, but we can conclude that an important part of the linear
aspect of the Arrhenius law is drasticaly modified by the detection system. 

\begin{figure}[htbp]
  \begin{center}
    \leavevmode
     \epsfxsize=15cm
     \epsfbox{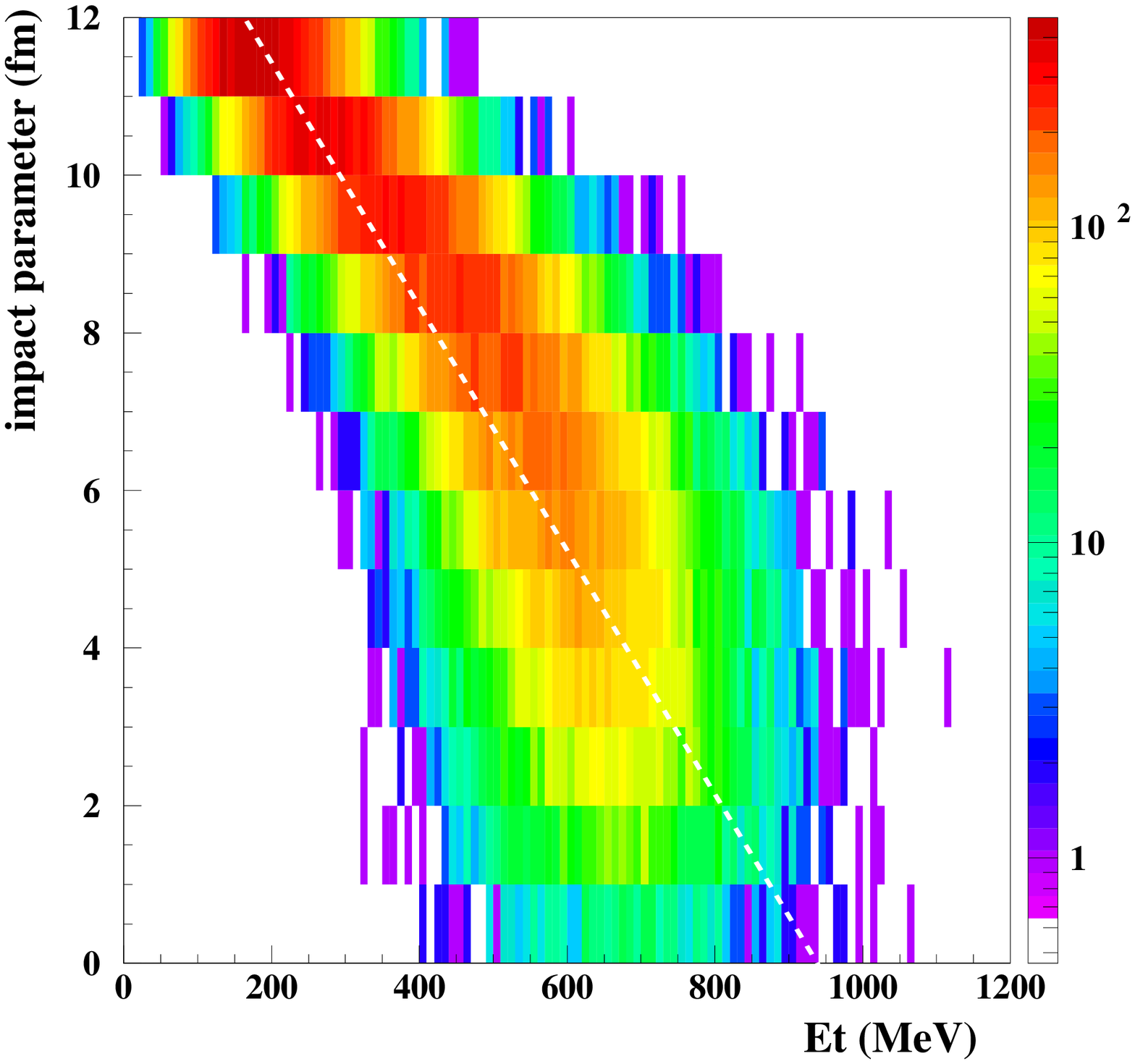}
      \caption{ }
    \label{fig:etvsb}
  \end{center}
\end{figure}

As there is a strong correlation between $E_{t}$ and $b$ (see
figure \ref{fig:etvsb}) it is usefull to perform the same
analysis for the impact parameter instead of $E_{t}$. From figure 3, we
extract on average, that $E_{T}\propto (\alpha b+b_{0}$) with $b_{0}=b_{max}$
and $\alpha <0$ then $1/\sqrt{E_{t}}\propto 1/\sqrt{b_{0}-\alpha b}$.

\begin{figure}[htbp]
  \begin{center}
    \leavevmode
     \epsfxsize=15cm
     \epsfbox{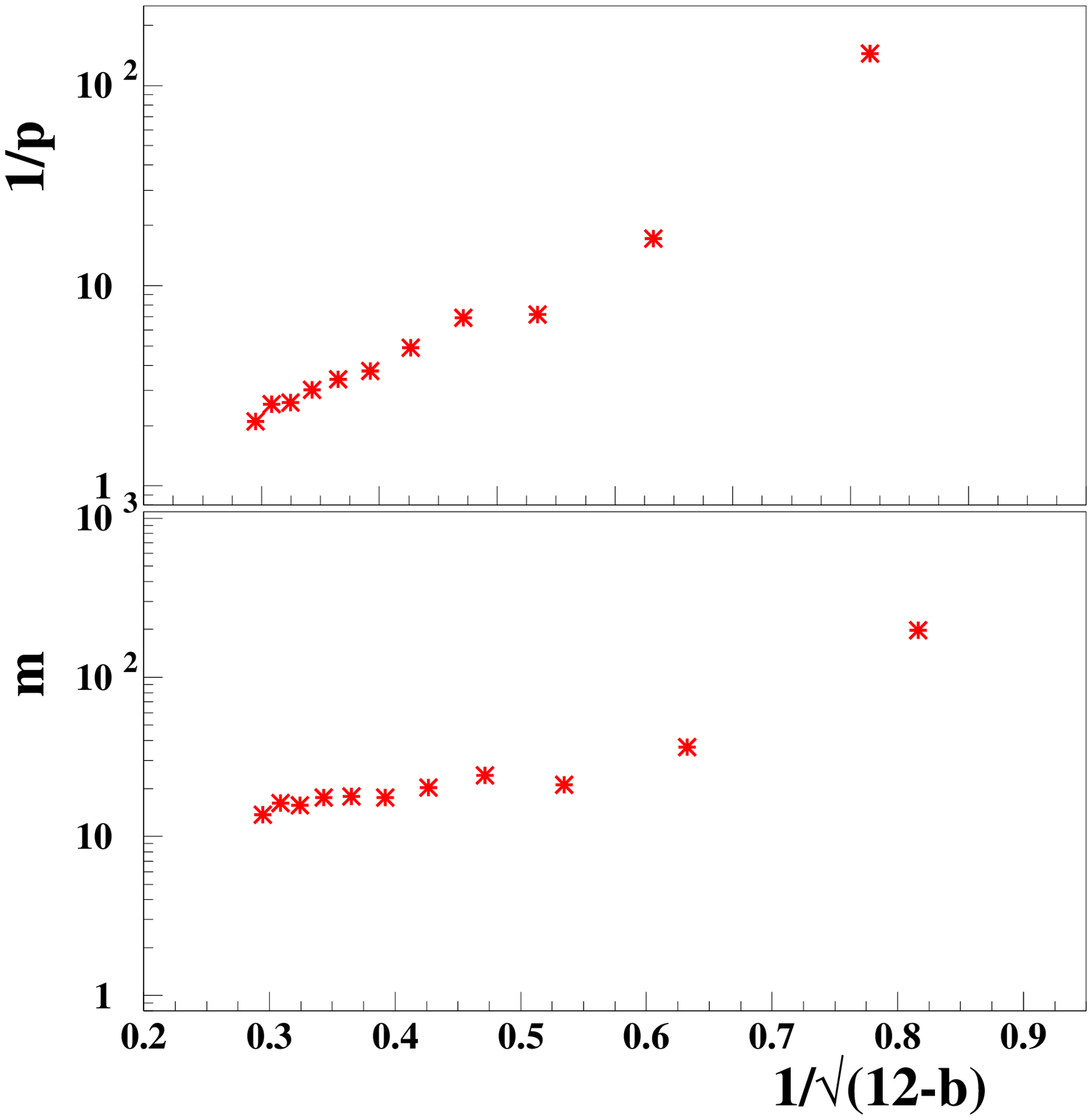}
      \caption{ }
    \label{fig:qmdb}
  \end{center}
\end{figure}

Figure \ref{fig:qmdb} shows the results, as a function of ${%
1/\sqrt{b_{0}-\alpha b}}$ with the same prescription as explained in table
\ref{tab:coupure}. We can see the same trends as in figure \ref{fig:qmdb}. In conclusion, the linear behavior can be due to geometrical
effects related to the impact parameter.

In this paper we demonstrated that the linear aspect of
Arrhenius plot in the multi-fragmentation regime depends strongly on the
acceptance of the detection system. A recent publication \cite{botvina2} shows also the effect of the detection system. The fact that, we can find the linear
behaviour with a non-thermal process could not be a justification of thermal
nature of multi-fragment emission process. Furthermore, this can be du to
the geometrical effect in the reaction including all the impact parameter.

Acknowledgement : we thank E.Plagnol for stimulating discussions. 

\clearpage

\begin{center}
Figure Caption
\end{center}

Figure \ref{fig:qmdf} : Top panel elementary probability $1/p$ versus $1/\sqrt{E_{t}}$,
bottom panel number of attempt $m$ versus $1/\sqrt{E_{t}}$. The plain line
correspond to the INDRA results for Xe + Sn at 50 A.MeV \cite{wieloch}, is
presented here like ''guide the eye'' because the x axis $E_{t}$ is not
totally reproduced by QMD code.

Figure \ref{fig:pm} : Right panel $1/p$ versus $1/\sqrt{E_{t}}$, left panel $m$ versus $1/\sqrt{E_{t}}$. The full circle represent the results for QMD with $ N_{IMF}=N_{frag}-QP-QT$. The open circle represente the results for QMD
with the prescription explain in table \ref{tab:coupure}.

Figure \ref{fig:etvsb} : Correlation between the impact parameter and the transverse energy in the QMD model for the reaction Xe+Sn at 50 A.MeV. Dashed line, the linear fit between $b$ versus $E_{t}$ used in figure \ref{fig:qmdb} see text.

Figure \ref{fig:qmdb} : Arrhenius plot for QMD events. The x axis is $1/\sqrt{b_{max}-b}$ (see text).



\end{document}